\documentclass[conference]{IEEEtran}
\IEEEoverridecommandlockouts
\usepackage{cite}
\usepackage{amsmath,amssymb,amsfonts}
\usepackage{algorithmic}
\usepackage{algorithm}
\usepackage{graphicx}
\usepackage{textcomp}
\usepackage{subcaption}
\usepackage{xcolor}
\usepackage{soul}
\sethlcolor{yellow}
\usepackage[normalem]{ulem}
\usepackage{enumitem}
\def\BibTeX{{\rm B\kern-.05em{\sc i\kern-.025em b}\kern-.08em
    T\kern-.1667em\lower.7ex\hbox{E}\kern-.125emX}}

\renewcommand{\baselinestretch}{0.98}

\begin{document}

\title{Joint SDN Synchronization and Controller Placement in Wireless Networks using Deep Reinforcement Learning
}

\author{\IEEEauthorblockN{Akrit Mudvari, Leandros Tassiulas}
\IEEEauthorblockA{ \textit{Yale University, New Haven, CT} \\
Email: akrit.mudvari@yale.edu, leandros.tassiulas@yale.edu}
\\[-6ex]
}

\maketitle

\begin{abstract}

Software Defined Networking has afforded numerous benefits to the network users but there are certain persisting issues with this technology, two of which are scalability and privacy. The natural solution to overcoming these limitations is a distributed SDN controller architecture where multiple controllers are deployed over the network, with each controller orchestrating a certain segment of the network. However, since the centralized control is the key attribute of SDN that allows it to be so beneficial, a centralized logical view of the network will have to be maintained by each of these controllers; this can be done through synchronization of the distributed controllers, where each controller communicates with the others to ensure that they remain informed about the entire network. There is however a network cost associated with constantly having to update each others about different aspects of the network, which will become a greater issue in dynamic wireless networks. To minimize this network cost, there is a need to consider not only when to get the update information from the neighboring controllers, but also where to dynamically place the controllers such that the network costs may be minimized. The placement should take into consideration both communication for synchronization among the distributed controllers and communication of the controllers with the network devices that they manage. In this work, we show that our multi-objective deep reinforcement learning-based method performs the best at achieving different application goals by developing policy for controller synchronization as well as placement, outperforming different other possible approaches, under a wide variety of network conditions. 

\end{abstract}

\vspace{-0.5em}
\section{Introduction}

One of the salient features of the next generation networks (including but not limited to wireless beyond-5G cellular networks) will be a multi-vendor multi-domain communication infrastructure where different entities will cooperatively orchestrate the network services for their clients \cite{neto2021seamless}. In cellular networks of the future for instance, such an effort is being facilitated with the research and development of off-the-shelf hardware and software deployments such as OpenAirInterface (OAI) \cite{oai-5g}, which is an open-source implementation spanning the full stack for the 5G (and other) cellular networks \cite{ghosh20195g}. Another example of networking paradigm that could become more proliferated and commercially viable in the future is federated cloud \cite{rochwerger2009reservoir}, where different cloud services can be accessed in a decentralized manner based on different bidding/management schemes. Such different networking infrastructures will communicate with each others, leading to a network of networks that will be different than the legacy deployments in that they will be orchestrated under multi-vendor SDN-driven methods. And as these networks grow, in terms of not just the number of the participating network entities, but also the number of vendors/controllers and the types of networks, we expect new challenges to limit or adversely effect the performance because of the limitations faced by the SDN technology.  

 SDN today is already a widely used network architecture that has been implemented across various types of network regimes including the large data centers. This is because of their capability in alleviating the limitations of the traditional communication architectures such as complex designs that require expertise for setup and reconfiguration, and costs associated with moving or adjusting hardware \cite{karakus2017survey}. A lot of these features of SDN are enabled by management of the control plane in a centralized and softwarized method \cite{kim2013improving}, but the same centralized approach also presents multiple limitation including scalability issues \cite{yeganeh2013scalability}: when a very large number of network devices have to be orchestrated by a single controller, it means that all the network entities will demand attention and resources from this single controller, including for monitoring network states, giving routing instructions, and many other tasks. Let us take for instance the task of populating the forwarding tables of the switches under an SDN controller: in absence of a resilient distributed protocol, failures of links will quickly translate to network-wide routing failure, which means that not only is it necessary to compute and update all the network decisions through this central entity, they have to be done quickly, at all times. And a failure of the controller spells disaster for the entire network \cite{oktian2017distributed}. 

\begin{figure}[t]
  \centerline{
    \includegraphics[width=0.49\textwidth]{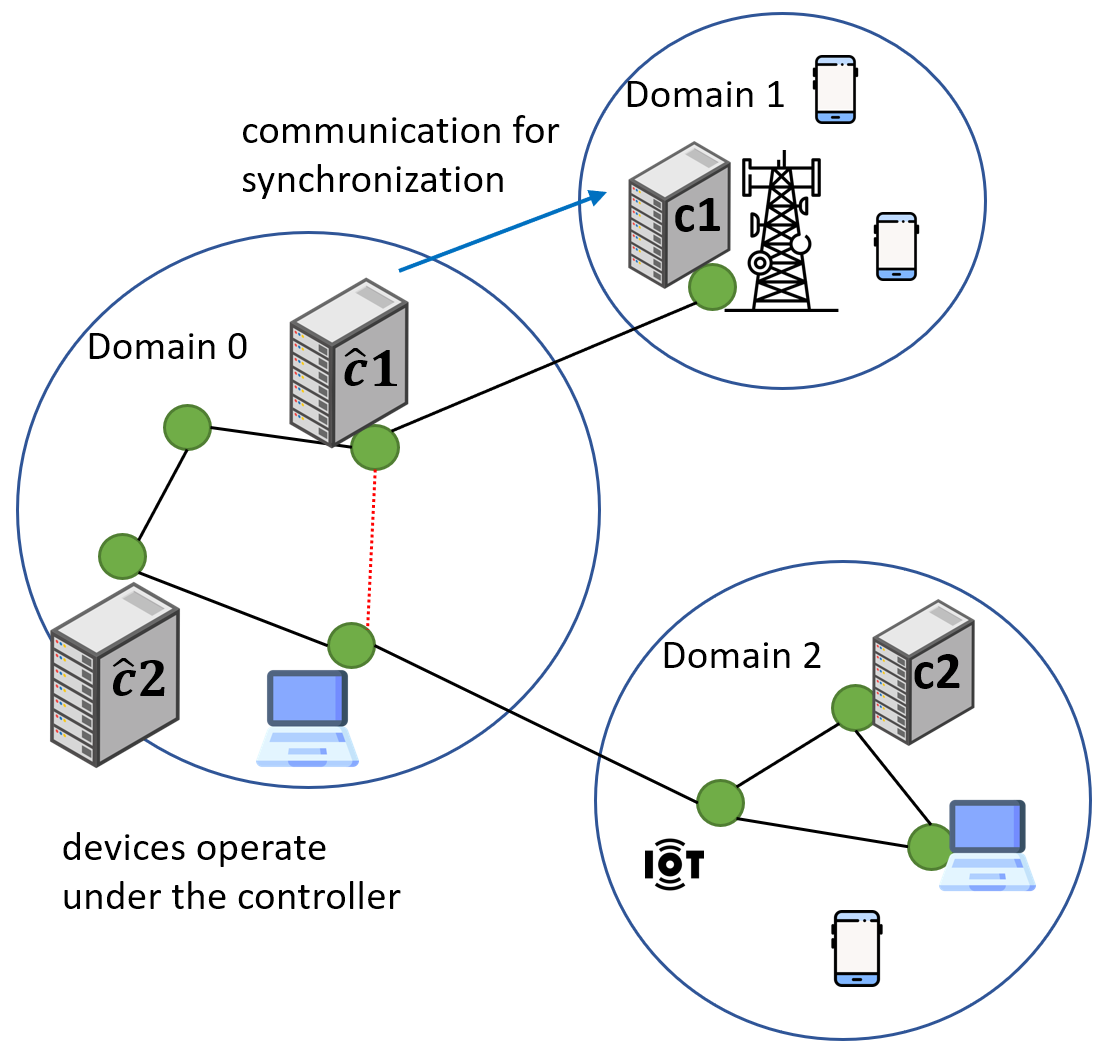}
  }
  \caption{Illustration of the distributed SDN controller paradigm in a dynamic network environment with distributed controllers.} 
  \label{fig:illustration}
\end{figure}

Solution to these problems pertaining to the centralized nature of the SDN controller comes in the form of distributed SDN architecture, where different physical controllers are used to orchestrate different subsections of the provided network \cite{yeganeh2013scalability}. This allows different controllers to manage a certain subsection of the network, which will greatly diminish the scalability issues, and ensure that failures are less globalized and solved quicker (these subsections are referred to as domains). The distributed approach also has the added benefit of allowing better privacy/security managements because all the information may not have to be (or must not be) pooled into a single controller, such as in situations exemplified by the aforementioned multi-domain multi-vendor networks; instead, the different controllers may have authority over different domains (we will refer to the subsection of a network that a distributed controller manages as its domain). However, it is very important for all the different controllers to have a global network state information of the entire network to make an informed decision, i.e., on "how to populate the routing tables of the devices the controller is responsible for such that the forwarded packets to any destination take the optimal route?". In wireless and dynamic networks, it becomes even more important to keep track of these information often since the network states tend to be volatile \cite{poularakis2019learning}. 

While a broad range of protocols/methods may be used for sending the synchronization messages amongst distributed controllers, the approach may broadly be classified into two categories: strongly consistent and eventually consistent \cite{botelho2016design}. The strongly consistent approach attempts to keep all the controllers synchronized at all times, including through protocols like RAFT \cite{ongaro2014search}.  For instance, OpenDaylight \cite{opendaylightWeb} and ONOS \cite{onosWeb}, which are some of the leading SDN implementations, implement RAFT, where each of the participant controllers have to agree on the same values of the global network state. With this approach, there is a higher chance of each of the controllers being synchronized at all times, but the communication cost of maintaining strong consistency is also high. On the other hand, with eventual consistency, we can allow temporary inconsistencies in the global network state as long as the effects aren't significant enough. And the benefit is that network costs associated with synchronization can be reduced. In \cite{zhang2019dq}, deep reinforcement learning  is used to achieve controller synchronization to improve inter-domain routing while achieving eventual consistency. In \cite{mudvari2023robust}, the synchronization policy is extended with transfer learning-based approach such that the policy remains efficient under changing controller environment. Eventual consistency is more desirable in the cases such as wireless networks and tactical networks \cite{phemius2016bringing}, where the network states change quite often but the communication infrastructures tend to have more limitations, since it is very important to be mindful of the excessive control plane communication between distributed controllers.

Under an SDN architecture, placement of controllers needs to considered in order to alleviate delays, specially control plane delays, in the network. For instance, K-means algorithm can be used to decide on an ideal placement of the controller, if the goal is to find a location such that the delay while communicating with the devices needs to be minimized \cite{gong2021social}. Such mechanism could also extend to a distributed controller environment, where multiple clusters can be obtained, and then controllers are assigned to each cluster such that the performance of the controllers can be improved, including the propagation latency \cite{liao2017density}. While some of the methods including the ones described above are static in nature, dynamic controller placement mechanisms can be considered to ensure that the placement decisions are sensitive to volatile and unpredictable network conditions \cite{dixit2013towards}. For instance, in \cite{bari2013dynamic} a dynamic controller placement strategy is developed such that the controllers are reassigned in response to the changing network conditions, with a goal being the reduction of communication overhead. 

Finally, it is necessary to consider the inevitable relationship between synchronization and controller placement. Take for instance figure \ref{fig:illustration} that exemplifies a dynamic wireless network with distributed controller deployment. Consider a situation when a link to controller $\hat{c}1$, red dotted line, breaks at some point in time. Then, it becomes beneficial to reassign controller to $\hat{c}2$, so that the average distance to the neighboring domains' controllers, $c1$ and $c2$, can remain relatively same. However, with eventual consistency-based synchronization policy in mind, it might be even more important to move closer to $c2$ since the topology it controls is much larger than the one under $c1$, and it might be more important for the controller in domain 0 to constantly remain updated about domain 2 under $c2$, so as to quickly respond to changes like link failures in the more communicated neighboring domain. Hence, an efficient policy would consider both synchronization and placement tasks together.

In this work, we uniquely consider both controller synchronization policy and controller placement policy at the same time, with the aim to maximize efficiency for both synchronization and placement goals, both of which depend on one another in a distributed control paradigm. Towards this goal: 

\begin{itemize}

\item We propose a novel approach that considers not just intra-domain (between controllers and switches) communication costs, but also inter-domain communication costs and synchronization decisions, towards dynamically deciding on synchronization and placement policies.

\item We propose a Double Deep Reinforcement Learning (DDRL) algorithm for optimally updating the network information among controllers in a dynamic environment, where the state of the network is dynamically changing and the policy needs to be adaptive for managing the global consistency under a strict communication budget. This is very important in wireless, cellular and other other networks where network state does not remain static.  

\item We show that our method produces state-of-the-art results for different types of applications under a dynamic network environment; this includes applications where the communication network features are shared among the domains (shortest path routing application) and where the end devices' properties are shared for decision making (load balancing computational task application). Furthermore, we show that this best performance is not only for a joint goal, but also for the synchronization and the placement goals individually.

\item We demonstrate the efficiency of our approach under different network environments and operator's decisions. 


\end{itemize}

\section{Problem Formulation} \label{section:formulation}

Towards creating a formulation for the problem we are trying to solve efficiently, which is to jointly optimize the synchronization and the placement of the controllers in a distributed SDN environment, consider the illustration in figure \ref{fig:illustration}. This is a communication system consisting of a certain number of domains, with each domain consisting of a certain number of network devices and a controller assigned to orchestrate the domain as well as maintain synchronized global state for the entire network by communicating with the other controllers of the neighboring domains. Note that while only one controller is assigned at a given time for a given domain, there are multiple possible locations for the controller placement within the domain, and this location can be changed dynamically over time to maintain efficiency. The formulation we will work on will be such that it works for any application as long as the reward is well-defined.

\subsection {Network model}

For a given domain, all the possible placement locations for the controller is denoted by a set $\hat{P}$, such that the selected controller at any time $t$ is given by $\hat{c}_t\in\hat{P}$, which we will interchangeably used as $\hat{c}$ when not referring to particular time step. Note that we will allow the policy to update after every time interval $\delta t$, such that the updates take place at time $t \in {1,2,3....}$.  The assigned controller $\hat{c}$ is then in a set $C$ of all assigned controllers serving under the global network, and so $ \hat{c}$ will have as its neighbors all the nodes $c \in C \char`\\ \hat{c} $, and $\hat{c}$ needs to remain in contact with all the neighbors as much as possible so that the most recent global network model can be maintained.

For any arbitrary application $a$, we will allocate a certain synchronization budget ${B_{sync}}^a_{\hat{c}}$, which signifies how often the controller $\hat{c}$ communicates with its neighbors. If this budget were to be exceeded and the synchronization were to happen more often, the network infrastructure would be overburdened with inter-controller control plane network traffic, so the goal of the synchronization task is to use this budget as a constraint and come up with the best synchronization policy that distributes the limited budget while providing the best outcome for a defined application $a$.  We will be allowed to take and update network information from ${B_{sync}}^a_{\hat{c}}$ neighbors at time $t$, which means that the information from the remaining neighbors will not be updated; the network information for those remaining domains will become $\delta t$ older. $ \hat{c}$'s policy for updating with a neighbor $c$ at time $t$ is given by $x_{\hat{c}c}(t) \in \{0,1\}$, where a value of 1 signifies updating will take place, and a value of 0 signifies that it will not, resulting in a temporary inconsistency. The following inequality defines the constraint to ensure that communication budget is not violated by any neighboring controller and any application, at a given time:

\begin{equation}
    \sum\limits_{c\in C \char`\\ \hat{c}}{x_{\hat{c}c}(t)} \leq {B_{sync}}^a_{\hat{c}}
\end{equation}

The placement budget ${B_{place}}^a_{\hat{c}}$  is always 1, which means that for a given domain in the network, only 1 controller will be used for orchestration at a time.

Throughout the learning process we will always allow the synchronization part of the policy to update every $t$, but we will allow the placement policy to update only after every $\tau t$ where $\tau \geq 1$. Since the placement decisions are made much less frequently then the synchronization decisions (to avoid controller migration costs), we could often have $\tau >> 1$. For instance, we observed that migration may take about 8 seconds for moving from one controller to another over a wireless network topology with an ONOS controller implementation; on the other hand, SDN synchronization rates may be in sub-second levels, depending on how fast the network features are evolving and how quickly applications need to respond to the changes. The implication of this design is that the synchronization actions (which define when the network state messages are shared) happen at a rate lower than the rate at which the placement decisions will be updated. 

Depending on the environment and to a certain extent the requirements of the application as well, we expect the allocation policy, inferred from the values of $x_{\hat{c}c}(t)$, to be optimal in different ways. For instance, if the task is shortest path routing and if the domain has a network of nodes that is very large and dense, it probably makes sense to allocate more of the budget to communicating with that particular domain, since the failure to maintain consistency will result in failure to communicate with more nodes. Since the policy is dependent on multiple aspects of the application and the environment, it becomes very difficult to formulate and predict the exact policy that will lead to the best allocation of the synchronization and the placement budgets. To overcome this complexity, we develop a DDRL (Double Deep Reinforcement Learning) model that learns an efficient allocation policy over a period of time.

\subsection {Multi-objective RL formulation for joint synchronization and routing}

Next we formulate the joint controller synchronization and placement problem as a Markov Decision Process (MDP). For the application-centric synchronization task, the first step is to choose a state space that is agnostic to the application and only depends on the core network structure. This state nonetheless needs to capture 'staleness' of the network state information in a dynamic environment, so it should represent "how long has it been since an update with a neighboring controller took place?". Hence, we will choose as the synchronization part of the state, $S_{sync}$, a vector of size $|C \char`\\ \hat{c}|$ with each entry corresponding to each neighbor $c$  representing the time steps since the last time an update with that neighbor took place in terms of $\delta t$. This will represent a sufficiently small state space that a learning agent will be able to handle in a sufficient time, towards developing an efficient yet dynamically adapting algorithm. For the placement task for the domain with controller $\hat{c}$, the state space $S_{place}$ will be a vector of size $\hat{P}$, with each value stating "how long the controller has been assigned this position in terms of $\delta t$?". A higher value signifies that the controller has not been "re-optimized" after placement for a longer time, and so $S_{place}$ is also an application-agnostic component like $S_{sync}$.

Similar to the state space, the action space will also consist of two components, $A_{sync}$ for synchronization and $A_{place}$ for placement task. $A_{sync}$ will be a vector of size $|C \char`\\ \hat{c} |$, with each entry corresponding to a neighboring controller or variable $x_{\hat{c}c}$ (decision by $\hat{c}$ to get update information from  the neighbor $c$). Hence there will be $B_{sync}$ 1's in the vector telling the controller to talk to those particular neighbors, with the rest of the values being 0. Similarly, $A_{place}$ will be a vector of size $|\hat{P}|$, with each entry corresponding to whether or not one of the possible placement locations are selected for the controller; this means one of the values in $A_{place}$ will be 1 commanding that the controller be placed there, and the rest will remain zeros.  

Up until now, our formulation allows for the MDP to be solved as a multi-agent reinforcement learning problem \cite{oroojlooy2022review} with different agents handling the synchronization and the placement tasks. However, as discussed earlier, the tasks and their rewards are not independent of each other, and as a result the agents working separately will demonstrate adversarial behavior, with each of them are trying to maximize their rewards while potentially harming the other's chance of reward maximization. For instance, the placement location may be chosen to minimize communication time solely based on inter-controller distance. But the synchronization agent may decide that certain neighbors are contacted a lot more, which means that those neighbors may need to be given extra weight while making the placement decision. On the other hand, by creating a single learning agent that combines the goals to create a multi-objective optimization problem \cite{mossalam2016multi}, we could remove the chances of adversarial behavior, including but not limited to an example scenario given above. With this in mind, we define the state space $S$ by combining the two state vectors: $S=(S_{sync},S_{place})$. And the action space $A$ is similarly created by combining the two action vectors, which the single learning agent will handle. For computational ease, we are concatenating the two vectors in case of space and action states to create the new vectors $S$ and $A$. As mentioned earlier in section \ref{section:ModelSetup}, the decisions for placement tasks are to take place every $\tau t$ as opposed to $t$ for synchronization. While our initial efforts to make the learning agent learn $\tau$ as a soft constraint was met with inefficiency (since learning $\tau$ becomes the third objective in a multi-objective formulation), we let the environment decide on $\tau$ (environment updates placement decision only at $t\bmod{} \tau =0$). With this value not known to the learning agent but also not a part of its learning objective, we see that this problem becomes a partially observable MDP, which as seen previously in \cite{hausknecht2015deep} and our own experiments, still works well when the learning agent is based on DRL (Deep Reinforcement Learning). 

Here, reward $R(S,A)$ is the reward after each step when for a given state $S$, an action in $A$ is performed. Just as in case of state and action, reward in this RL formulation consists of two components: $R_{sync}$ for synchronization and $R_{place}$ for placement. The reward for synchronization tasks $R_{sync}$ is application centric, and will depend on the goals of the synchronization policy. For instance, for a load balancing task, $R_{sync}$ could be the difference in computation latency caused when the offloading task is send to a less efficient server destination because the domain is unaware of where the best server locations are. Hence, the network engineer could select this reward in accordance with the application goals (We will discuss the specifics further in section \ref{section:discussion}). Reward for placement tasks $R_{place}$ represents the reward obtained through optimal placement of the controllers. In a centralized SDN architecture, the goal of control placement is based on how well the controller will be able to orchestrate the entire network in an optimal way, i.e., by minimizing the control plane delays \cite{gong2021social}. However, since we are dealing with a distributed SDN environment, the placement of the controller will also have to take into consideration the control plane traffic for inter-controller communications , including the ones being used for controller synchronization. 

We selected placement reward as shown in equation \ref{eqn:placeReward}:

\begin{equation}
    \label{eqn:placeReward}
    \begin{split}
    R_{place}(S,A) = \left(\sum_{e \in \hat{E}}{\delta(\hat{c},e)}\right)^{-1}
     +  
    \mu \left(\sum_{c \in C \char`\\ \hat{c}}{\delta(\hat{c},c)}\right)^{-1}
    \end{split}
\end{equation}

where $\delta(a,b)$ would represent the communication delay between two nodes $a$ and $b$ in the network, and we also introduce variable $e \in \hat{E}$ to represent each switch $e$ in the subdomain controlled by controller $\hat{c}$. In equation \ref{eqn:placeReward}, the placement reward is the sum of two component, where the first part is the inverse of the sum of delays faced by the controller $\hat{c}$ while communicating with each of the device $e$ that it controls/ exists in its domain. The second part is the inverse of the sum of delays faced by the controller $\hat{c}$ while communicating with each of the neighboring controllers. So a higher reward would be achieved by placing the controller as close as possible to neighboring controllers as well as to the device within the domain. Furthermore, by considering, at time $t$, only those neighboring controllers to which we are trying to synchronize at time $t$ as defined by action $A_{sync}$, we further optimize the placement policy and synchronization policy together. $\mu$ is a parameter that decides the relative importance of the control plane communication within the subdomain versus among the neighboring controllers, with higher $\mu$ causing the reward to focus more on inter-domain communications. The total reward is then defined in equation \ref{eqn:totalReward}: 

\begin{equation}
    \label{eqn:totalReward}
    R = R_{sync}(\alpha +R_{place})
\end{equation}

where the parameter $\alpha$ is used to signify the relative importance of the placement goal in comparison with the synchronization goal. Different synchronization goals may be adopted for learning under our method, but this approach was observed to work well for different goals as seen in \ref{section:discussion} section. Thus we have $M=(S,A,R)$, and the goal is to find the best policy $\pi*$ that maximizes the sum of reward over a period of time $T$.  

Our method is based on Q learning \cite{watkins1992q}, where the Q function values are allocated for the explored state-action pairs using Deep Neural Networks (DNNs); to be more specific we implement a double DQRL approach \cite{mnih2015human}, where a lagging DNN model called target network learns slower than the main network, and in the process we avoid over-fitting. The DNN used in the implementation is a multiplayer perceptron \cite{kumar2011multilayer}, with the state values as the input, and the action values as the output, with each output node in the DNN corresponding to a unique possible action. The selected action after each DNN inference is the action node with the highest activation value, which we consider to be the $Q$ function output for the given state-action pair. The DNN is implemented using Pytorch library \cite{paszke2019pytorch}, which is an efficient tool for implementing neural network training and inference. And the optimizer chosen for learning is Adam \cite{kingma2014adam}. The DDRL scheduler developed as described is shown as Algorithm \ref{algorithm: joint_algorithm}.

\begin{algorithm}[b!]
    \caption{DDRL scheduler for joint synchronisation and placement}
    \label{algorithm: joint_algorithm}
    
    \textbf{Input:} parameters for initializing the NNs (NNs will start with randomized parameters $\theta$), learning rate $\alpha$, batch size $b$,  decay rate $\epsilon$,  soft update rate for target network $\kappa$, reward parameters $\mu$ and $\alpha$  \\
    \textbf{Output:} learned parameters $\theta$ for DNN \\
    \begin{algorithmic}
    \STATE Initialize:  Main and target NN with parameters $\theta$ (randomized) 
    \STATE Initialize: matrix $D$ for $(s,a,r,\phi)$ tuples to be stored
    \FOR{episodes $E = 1,....,M$}
        \STATE Set: exploration probability $P_{\epsilon} = \frac{1} {(1 +  E / \epsilon)}$
        \STATE Set: initial state $s_0$ to zero vector, $t=0$
        \WHILE{$t\leq T$}
            
            \STATE with probability $P_{\epsilon}$ select random action $a_t$ 
            \STATE otherwise select $a(t) = argmax_{a} Q(s(t),a;\theta$)
            \STATE From network environment obtain the rewards $r_{sync}(t)$ and $r_{place}$ according to $(s(t),a(t))$, and the observation $\phi(t)=s(t+1)$
            \STATE $r_{place}$ given by equation \ref{eqn:placeReward}
            \STATE calculate $r(t) = r_{sync}(t) (\alpha + r_{place}(t))$ 
            \STATE Store $(s(t),a(t),r(t),\phi(t))$ in $D$
            \STATE Sample random mini-batch of size $b$ from $D$, each entry denoted by $(s_j,a_j,r_j,\phi_j), j=1,...,b$
            \FOR {$j=1,...,b$}
                \STATE $y_j = r_j + \gamma Q_{\theta}(S_j,\underset{a \in A}{\mathrm{argmax}} Q_{\hat{\theta}}(S_j,A_j)$  
                \STATE perform gradient descent step on $(y_j - Q(s_j,a_j;\theta))^2 $ with respect to $\theta$, to obtain $\nabla_{\theta}L(\theta)$
            \ENDFOR
            \STATE $\theta \leftarrow \theta - \alpha \nabla_{\theta}L(\theta)$  
            \STATE $\hat{\theta} \leftarrow \kappa \theta + (1-\kappa)\hat{\theta} $
        \ENDWHILE
    \ENDFOR
    \end{algorithmic}
\end{algorithm}

\subsection {Application Verification}

We verify the efficiency of our method on multiple test applications to ensure that the method works for different application centric goals. The first application will be Shortest Path Routing (SPR), where the controller will have to remain updated about the global network state so that the most up-to-date and globally optimal routing decisions can be calculated by the controller, and provided to the participating nodes for SPR. Unlike in context of distributed protocols (i.e., OSPF), the controllers must be the ones to ensure that the forwarding tables of the switches are correctly populated, so if the most recent information about the neighboring domains is not available, the paths calculated may not be the most efficient ones. Another application considered is Load Balancing (LB), which refers to efficiently designating incoming traffic to the optimal group of potential servers; SDNs have been shown to effectively perform LB tasks \cite{qilin2015load}, more so than the traditional methods. While SPR focuses on the links states, LB is primarily dependent on the end-node capacities, which means we investigate over different types potential applications.

\section {Evaluation and Discussion}
\label{section:discussion}

We now show that our joint scheduler outperforms other methods like round robin and randomization methods. In section \ref{section:ModelSetup}, we discuss the 
environment and parameters used for evaluation, and then in section \ref{section:EmuResults}, we show the improvements in both synchronization as well as placement decisions when our DRL-based method is used. And finally in section \ref{section:envronment} we consider various environment factors and network operation decisions towards showing versatility of our implementation. 

\subsection{Network model setup} \label{section:ModelSetup} 

For each of the experiments, we generate randomized and dynamic topologies that evolve over time. We deploy certain number of controllers, with each of them randomly assigned a domain in the network, consisting of a random subset of interconnected switches. The number of controllers is initially selected as 7, with each controller assigned a randomized domain with 3 to 15 switches. The synchronization budget is predetermined, i.e., 28 percent. We allow the network to change gradually, with the links randomly going up or down, and the server capacities evolving. The learning rate for the learning agent is set at 0.01, the batch size is 256, the buffer for storing $(s(t),a(t),r(t),\phi(t)$ is maintained at 40,000, the epsilon decay denominator $\epsilon$ is 10, and the discount rate $\gamma$ is set to 0.1 (for LB) and 0.4 (or SPR). The reward parameter $\alpha$ (signifies relative importance of placement reward) is set to 2 (for SPR) and 1 (for LB). The relative update rate of placement problem $\tau$ is tested between randomized range of 10 to 100 as this is left to the environment/network operator. These choices aside, we generally observed that the learning methods worked well for a wide range of parameters selections, which we concluded after hyerparameter searching. While the placement reward is a constant expression (as given by equation \ref{eqn:placeReward} and explained earlier) throughout different cases, the synchronization rewards are a function of the application, and is defined in a way that best-represents the goals of that particular application. 

\begin{figure}[t]
  \centerline{
    \includegraphics[width=0.425\textwidth]{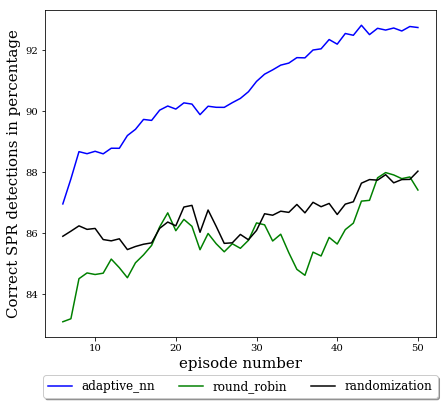}
  }
  \caption{Evaluation Results for SP synchronization} 
  \label{fig:sp_percent}
\end{figure}

\begin{figure}[b]
  \centerline{
    \includegraphics[width=0.375\textwidth]{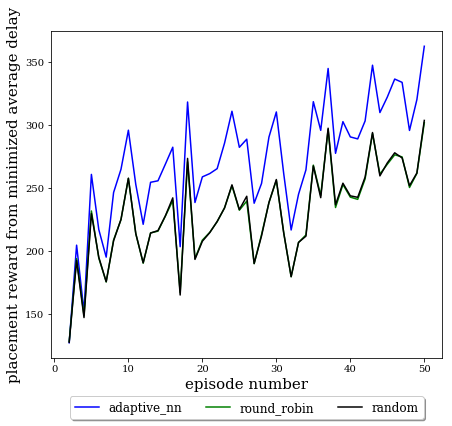}
  }
  \caption{Evaluation Results for SP placement} 
  \label{fig:sp_place}
\end{figure}

\begin{figure}[t]
  \centerline{
    \includegraphics[width=0.425\textwidth]{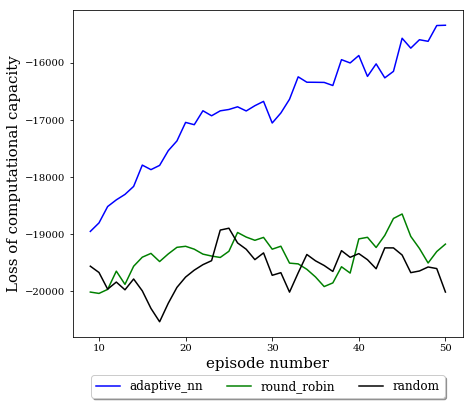}
  }
  \caption{Evaluation Result for LB synchronization} 
  \label{fig:lb_perform_diff}
\end{figure}

\begin{figure}[b]
  \centerline{
    \includegraphics[width=0.375\textwidth]{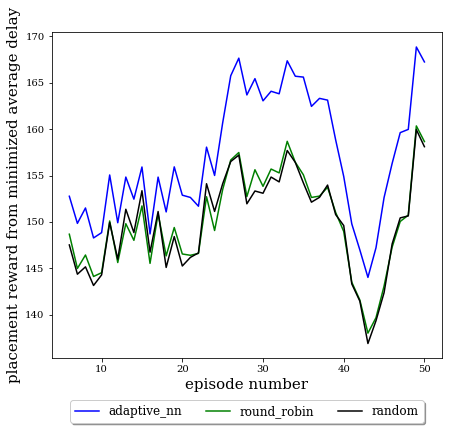}
  }
  \caption{Evaluation Result for LB placement} 
  \label{fig:lb_place}
\end{figure}

\subsection{Evaluation Results} \label{section:EmuResults}

While the learning agent was trying to maximize the overall goal as seen in equation \ref{eqn:totalReward}, from the application prospective this value is a combination of two separate and important practical goals: one ideal and dynamic placement of SDN controllers, and two synchronization goal as defined by the application. Next we try to show that both of these objectives were efficiently met for different application cases, while comparing against a randomization approach where resources are randomly allocated, and a round robin approach where resources are allocated in a round-robin manner.   

\textbf{Evaluation of SPR}: As seen in figure \ref{fig:sp_percent}, our method was shown to outperform both randomization and round robin approaches for better-predicting the shortest paths across the network (higher value signifies that larger percentage of the shortest paths were accurately discerned under the partial network state information). The performance for our method, adaptive\_nn (DDRL), was $6.4$ percent higher than round robin approach and $5.6$ percent higher than randomization approach. While initially the reward for adaptive\_nn (DDRL) evaluation was given by $R_{sync} = \sum_{t}k.detect(t)$ where $detect(t)$ is sum of all correct detection of shortest path in each time step $t$ throughout an episode, we analyse by looking at what percent of the shortest paths in the network were correctly identified. As seen in figure \ref{fig:sp_place}, we could also observe that the total reward obtained by our method was highest for the controller placement task as well, outperforming round robin approach by $20.2$ percent and randomization method by $20.0$ percent; this placement reward (for any application including SPR and LB) is defined by equation \ref{eqn:placeReward}.

\textbf{Evaluation of LB}: Similarly, we can see in figure \ref{fig:lb_perform_diff} that our approach adaptive\_nn (DDRL) is better than randomization and round robin approaches at predicting the best load balancing policy, i.e., at determining the allocation of tasks to the best server candidate. We see that our DRL-based method outperforms other approaches at all times, as signified by higher reward values; here the reward for each step (and hence the performance metric in figure \ref{fig:lb_perform_diff})  measures the difference between the assigned server capability of the actual best server location and the perceived server location under the synchronization policy. Note that if all neighboring domains were perfectly synchronized, this difference would not exist, but it does under the limited budget. Our policy outperforms randomization approach by $16.8$ percent and round robin approach by $18.1$ percent. As seen in figure \ref{fig:lb_place}, we could also observe that the total reward obtained by our method was highest for the controller placement task as well, outperforming round robin approach by $18.6$ percent and randomization method by $18.7$ percent.

\subsection{Analysis of environment factors} \label{section:envronment}

\begin{figure}[t]
  \centerline{
    \includegraphics[width=0.425\textwidth]{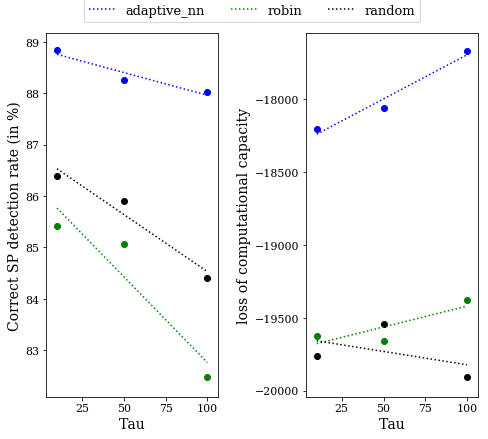}
  }
  \caption{Left: Correct SP detection Rate for different values of $\tau$. Right: Loss of Computational Capacity for different values of $\tau$.} 
  \label{fig:tauvals}
\end{figure}


\begin{figure}[t]
  \centerline{
    \includegraphics[width=0.425\textwidth]{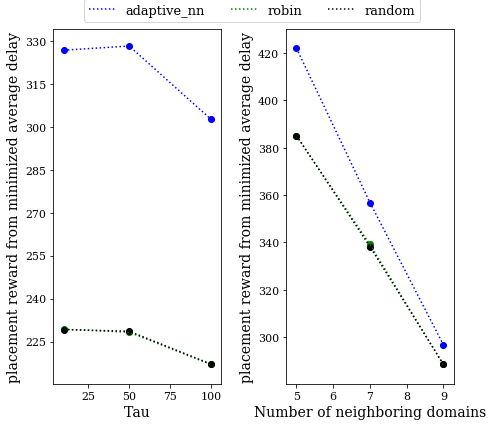}
  }
  \caption{Left: for SP application, placement rewards for different values of Tau($\tau$). Right: for LB application, placement rewards for networks with different number of SDN controllers to interact with.} 
  \label{fig:placeAnalysis}
\end{figure}

\begin{figure}[b]
  \centerline{
    \includegraphics[width=0.425\textwidth]{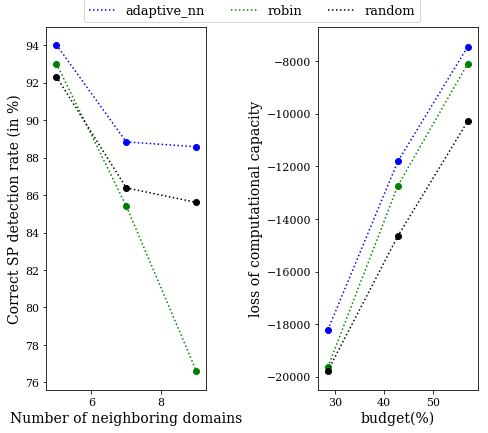}
  }
  \caption{Left: Loss of Computational Capacity for different budget availability. Right:   Correct SP detection Rate for networks with different number of SDN controllers to interact with} 
  \label{fig:syn_cbudnumC}
\end{figure}

\begin{figure}[t]
  \centerline{
    \includegraphics[width=0.425\textwidth]{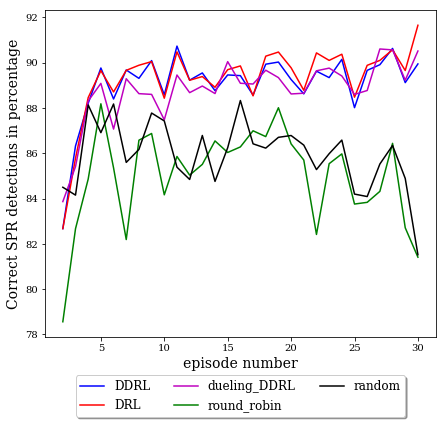}
  }
  \caption{Correct shortest path detection rate (SP application) for different methods analysed} 
  \label{fig:methods_sync}
\end{figure}

\begin{figure}[b]
  \centerline{
    \includegraphics[width=0.375\textwidth]{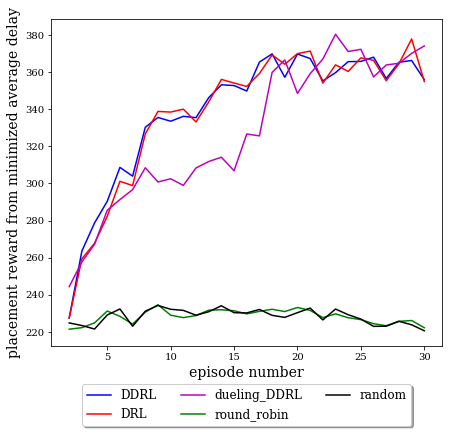}
  }
  \caption{Placement reward (SP application) for different methods analysed} 
  \label{fig:methods_place}
\end{figure}

In this section, we consider different network conditions, operator decisions, and other aspects such as different reinforcement learning methods that helped us arrive at an approach that is practically useful and widely feasible for the distributed SDN controller implementation. Note that a lot of hyperparameters that we discussed earlier held for these various conditions, which means the implementation does not require significant hyperparameter tuning across changing network conditions. Nonetheless, further tuning could be implemented for some marginal benefits. Please note that as mentioned earlier, our DDRL method is called adaptive\_nn in the figures discussed.

\textbf{Across different $\mathbf{\tau}$ values}: The decision of "how often should the placement decisions be updated?" relies on the network operator who might want to balance the cost of controller migration with an efficient placement and synchronization policy in a dynamic network environment. So observations were made to see how well the learning agent performs when the frequency of placement updates, denoted by $\tau$, is varied. We could see our method outperforming randomization and round robin approach for a range of values of $\tau$ as demonstrated in figure \ref{fig:tauvals}(left) for SP application, and figure \ref{fig:tauvals}(right) for LB application. We also show that our approach outperforms the other methods for finding the best placement location (i.e., lower average delay for communications as shown by higher placement reward) across different values of $\tau$ as shown in figure \ref{fig:placeAnalysis}(left). 

\textbf{Across different synchronization budgets and network structures}: Depending on factors such as data and control plane traffics, different synchronization budgets may be ideal to a  distributed controller environment; so in this section we show that for different budget levels, our method will show greater efficiency over methods like round robin and randomization. Similarly, it is also important that the method works across different types of network environment, so it is shown that our method performs better for different network environments with different number of neighboring controllers in the distributed SDN paradigm. In figure \ref{fig:syn_cbudnumC}(left) it is shown that our method outperforms other approaches for shortest path routing application for different budget levels. And in figure \ref{fig:syn_cbudnumC}(right), we show that our method does a better job at reducing the loss of computational capacity for the load balancing application across different number of SDN controllers in the distributed controller environment. In figure \ref{fig:placeAnalysis}(right), it is similarly shown that for the LB application, we observe that our method outperforms across different number of distributed SDN controllers at minimizing the average delay (i.e., in obtaining highest placement reward). We show our analysis for certain methods for sake of brevity for the readers, but these trends could be observed across different applications and factors including budget and number of distributed controllers.  


\textbf{Analysis of the potential methods}: In this subsection, we discuss how we explored different reinforcement learning methods to decide on the best RL approach. In figure \ref{fig:methods_sync} and \ref{fig:methods_place}, we observe the relative performance of different value-based RL methods towards obtaining the optimal reward for an application (SPR) as well as for an optimal placement decision respectively. We observed that while the different value-based RL methods outperformed round robin and randomization approaches, these methods had a very close performance, with the rewards tending to stay within 1 percent of each others when summed over different episodes. And the standard deviation was also fairly similar, denoting similar robustness in performance since this means the reward is less likely to fluctuate over time. DDRL appeared to be a method which had marginally better performance, resulting in our choice, but it could be observed that DRL or dueling DDRL \cite{wang2016dueling} methods could also perform quite well. Our MDP formulation was observed to lend itself well to different value-based RL methods; On top of that it has been observed that, for such discrete optimization problems where more practically frequent scenarios are explored more often, value-based RL methods do a better job, and similar observation has been made in \cite{zhang2019dq}. Furthermore, we have a model where it is quite important to avoid local minima, which the policy-based RL methods tend to be more vulnerable to; these observations and experiments led to our final decision of using the DDRL approach. 


\section{conclusion}

In this work, we explored a novel method for jointly optimizing SDN controller synchronization and placement tasks for dynamic, wireless network paradigms, taking into consideration the fact that the controller placement objective in a distributed SDN environment must also take into consideration the controller synchronization objective, and vice versa. We took into consideration the dynamic, ad-hoc nature of the network topologies where nodes and links in the topologies are randomly changing, and towards this goal, we developed and implemented a double deep reinforcement learning approach and formulated a suitable multi-objective goal. We showed that our method efficiently develops synchronization as well as placement policies for a range of applications such as shortest path routing and load balancing. 



\bibliographystyle{./bibliography/IEEEtran}
\bibliography{./bibliography/main}

\end{document}